\documentclass[aps,prl,reprint,a4paper,superscriptaddress,numbers,longbibliography,showpacs,showkeys]{revtex4-1}

\usepackage{amssymb}
\usepackage{amsmath}
\usepackage{bm}

\usepackage[pdftex]{hyperref,graphicx}
\usepackage[english]{babel}
\usepackage[utf8]{inputenc}
\usepackage[T1]{fontenc}
\usepackage{xspace}

\usepackage[separate-uncertainty=false]{siunitx}

\makeatletter
\newcommand{\longdash}[1]{{\emph{#1}}.\textemdash}
\renewcommand\section{\@startsection {section}{1}{9.24774pt}                                   {1pt}                                   {0em}                                   {\raggedright\longdash}}
\DeclareRobustCommand{\@seccntformat}[1]{  \def\temp@@a{#1}  \def\temp@@b{section}  \ifx\temp@@a\temp@@b
  \csname the#1\endcsname .\quad  \else
  \csname the#1\endcsname\quad  \fi
}
\makeatother

\newcommand{\figref}[2]{\hyperref[#1]{\getrefnumber{#1}(#2)}}

\addto\captionsenglish{}
\addto\captionsenglish{}

\hypersetup{
    pdftitle = {Spectral flow and global topology of the Hofstadter butterfly},
    pdfauthor = {János K. Asbóth, Andrea Alberti},
        colorlinks=true,linkcolor=blue,citecolor=blue,filecolor=blue,urlcolor=blue
  }

\newcommand{\abs}[1]{\left|{#1}\right|}

\newcommand{\Econt}{\tilde{E}}

\newcommand{\HH}{\hat{H}}

\newcommand{\UU}{\hat{U}} 
\newcommand{\Heff}{\HH_\textrm{eff}} 

\newcommand{\encapsulateMath}[1]{\raisebox{0pt}[0pt][0pt]{#1}}

\newcommand{\Upper}{\text{up}}
\newcommand{\Lower}{\text{low}}

\newcommand{\Egap}{\tilde{E}}

\newcommand{\Eupper}{E^{\Upper}}

\newcommand{\tr}{\mathrm{tr}\,}

\newcommand{\topind}{\nu}

\renewcommand\textemdash{\leavevmode\unskip\kern0.8pt\rule[0.19\baselineskip]{8pt}{0.4pt}\kern1pt\ignorespaces}

\begin{document}

\title{Spectral flow and global topology of the
  Hofstadter butterfly}
\author{J{\'a}nos K. Asb{\'o}th}
\affiliation{Institute for Solid State Physics and Optics, Wigner Research Centre for Physics, Hungarian Academy of Sciences, H-1525 Budapest P.O. Box 49, Hungary}
\author{Andrea Alberti}
\affiliation{Institut für Angewandte Physik, Universität Bonn, Wegelerstr.~8, D-53115 Bonn, Germany}
\date{Summer 2016}
\begin{abstract}
We study the relation between the global topology of the Hofstadter butterfly of a multiband insulator and the topological invariants of the underlying Hamiltonian.
The global topology of the butterfly, i.e., the displacement of the
energy gaps as the magnetic field is varied by one flux quantum, is
determined by the spectral flow of energy eigenstates crossing gaps as
the field is tuned.
We find that for each gap this spectral flow is equal to the
topological invariant of the gap, i.e., the net number of edge
modes traversing the gap.
For periodically driven systems, our results apply to the spectrum of
quasienergies.
In this case, the spectral flow of the sum of all the quasienergies
gives directly the Rudner invariant.
\end{abstract}
\maketitle

\begin{figure}[b] \includegraphics[width=\columnwidth]{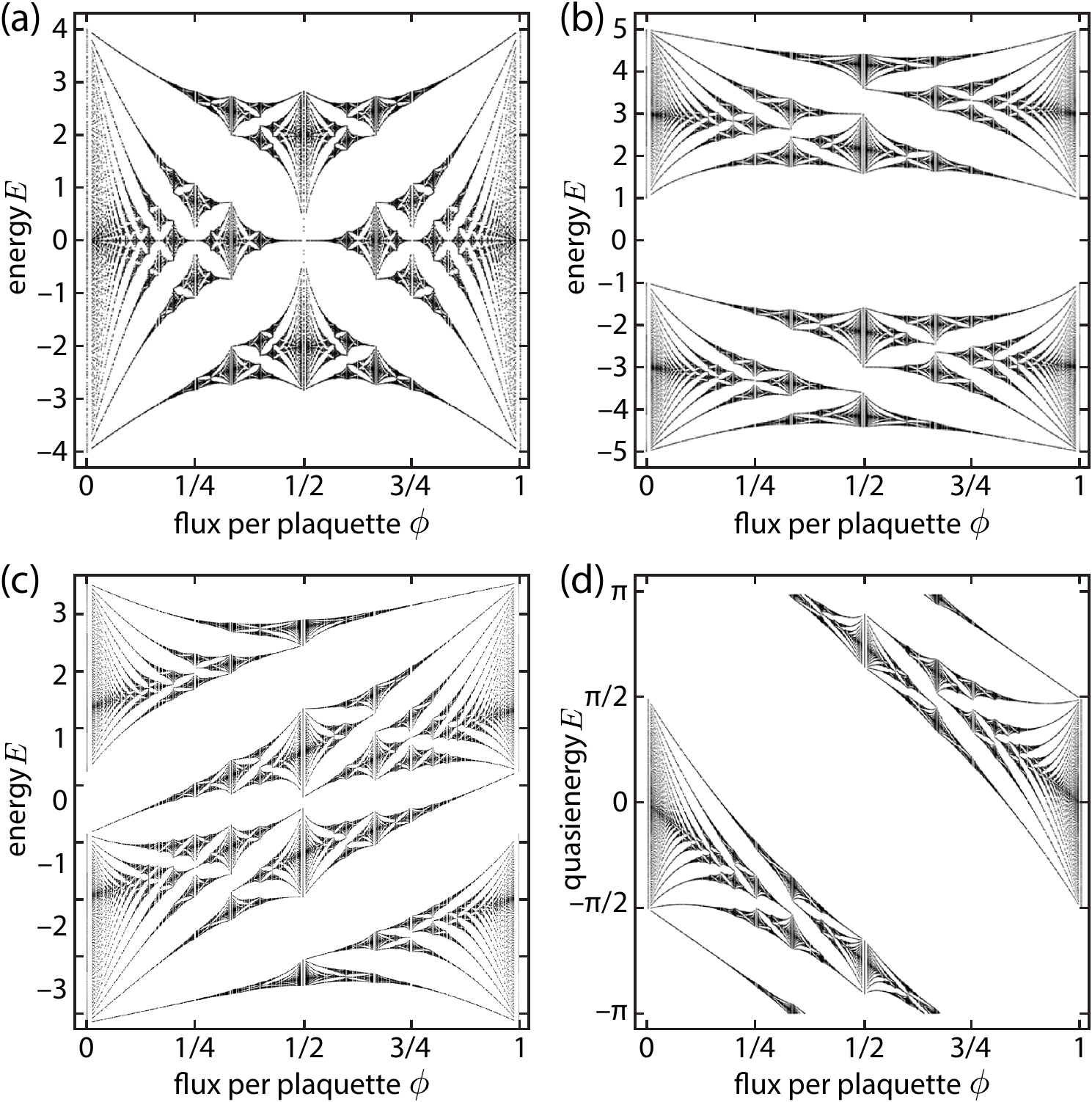}
\caption{\label{fig:four_examples} Examples of Hofstadter butterfly
  spectra for: (a) a simple charged particle on an infinite square lattice (Harper
  model), (b) a topologically trivial multiband insulator (Qi-Wu-Zhang
  model with $u=3$ of Ref.~\onlinecite{asboth2016book}), (c) a
  topological multiband insulator (same Qi-Wu-Zhang model with $u=1$),
  (d) a topological Floquet insulator (Rudner model of
  Ref.~\onlinecite{rudner2013anomalous}).\vspace{-4mm} }
\end{figure}

The Hofstadter butterfly is the self-similar structure of subgaps in
the energy spectrum of a charged particle hopping on a two-dimensional
lattice, as a function of a perpendicular magnetic field.
Its fractal structure becomes apparent when the magnetic flux, $\Phi$,
on each pla\-quette is a sizable fraction of the magnetic flux
quantum, $\Phi_0=h/Q$, i.e., the normalized flux $\phi=\Phi/\Phi_0$ is
of the order of 1 ($Q$ is the charge and $h$ is the Planck constant).
This is shown in Fig.~\figref{fig:four_examples}{a}.
Since it was first numerically computed \cite{hofstadter1976energy},
the Hofstadter butterfly has played an instrumental role in
understanding the quantum Hall
effect~\cite{thouless1982quantized,streda1982theory}, it has made
connections between number theory and physics
\cite{wannier1978result,MacDonald:1983}, and it has inspired numerous
other works (see, e.g., Ref.~\citenum{Satija:2016}).
Observation of the butterfly using traditional solid-state materials
would require prohibitively strong magnetic fields
(thousands of teslas).
Alternative approaches focus on enlarging the plaquette size using
superlattices, or substituting the magnetic field with a synthetic implementation of a vector potential (e.g.,
by rotation~\cite{umucalilar2008trapped} or laser-assisted
tunneling~\cite{jaksch2003creation,Gerbier:2010}).
There is recently a renewed interest in this problem because after so
many years both approaches have come close to ``netting'' the
Hofstadter butterfly, using heterostructure superlattices
\cite{Schloesser:1996,Albrecht:2001,Geisler:2004} or moiré
superlattices made of graphene on a
substrate \cite{dean2013hofstadter}, and using ultracold atoms in
``shaken'' optical
lattices~\cite{miyake2013realizing,aidelsburger2013realization}.

{\setlength{\parskip}{1pt plus 0pt minus 0pt}
The Hofstadter butterfly is known to be periodic: 
the spectrum is invariant under a shift of
$\phi$ by 1.
This also applies to multiband insulators, e.g., if the particle has
several internal states \footnote{Assuming we have a simple lattice,
  and the magnetic field couples only to the motion via Peierls
  phases, and not to the internal states.}.
In such a multiband Hofstadter butterfly, the periodicity is trivially obeyed if each band develops its own set of minigaps, as shown with an example in Fig.~\figref{fig:four_examples}{b}.
However, there exist more ways in which this constraint can be obeyed: bands can also flow into each other as $\phi$ is tuned from 0 to 1, as shown in Fig.~\figref{fig:four_examples}{c}.
We call this pattern of bands flowing into each other the \emph{global topology} of the Hofstadter butterfly.
An even wider variety of nontrivial global
  topologies  can
occur in a periodically driven system (Floquet system), where quasienergy takes the place of energy
\textemdash much like quasimomentum takes the place of momentum in a
lattice system.
In this case, bands can even \emph{wind} in quasienergy as shown in Fig.~\figref{fig:four_examples}{d}.
}

In this Letter, we establish a connection between the global topology
of a multiband Hofstadter butterfly and the topological invariants of
the underlying Hamiltonians.
For periodically driven systems, in particular, we find that the
winding of the Hofstadter butterfly is determined by the Rudner
invariant~\cite{rudner2013anomalous}
of the drive.
We give a direct formula for this invariant in terms of the sum of
quasienergy eigenvalues.

\begin{figure}[b]
\includegraphics[width=\columnwidth]{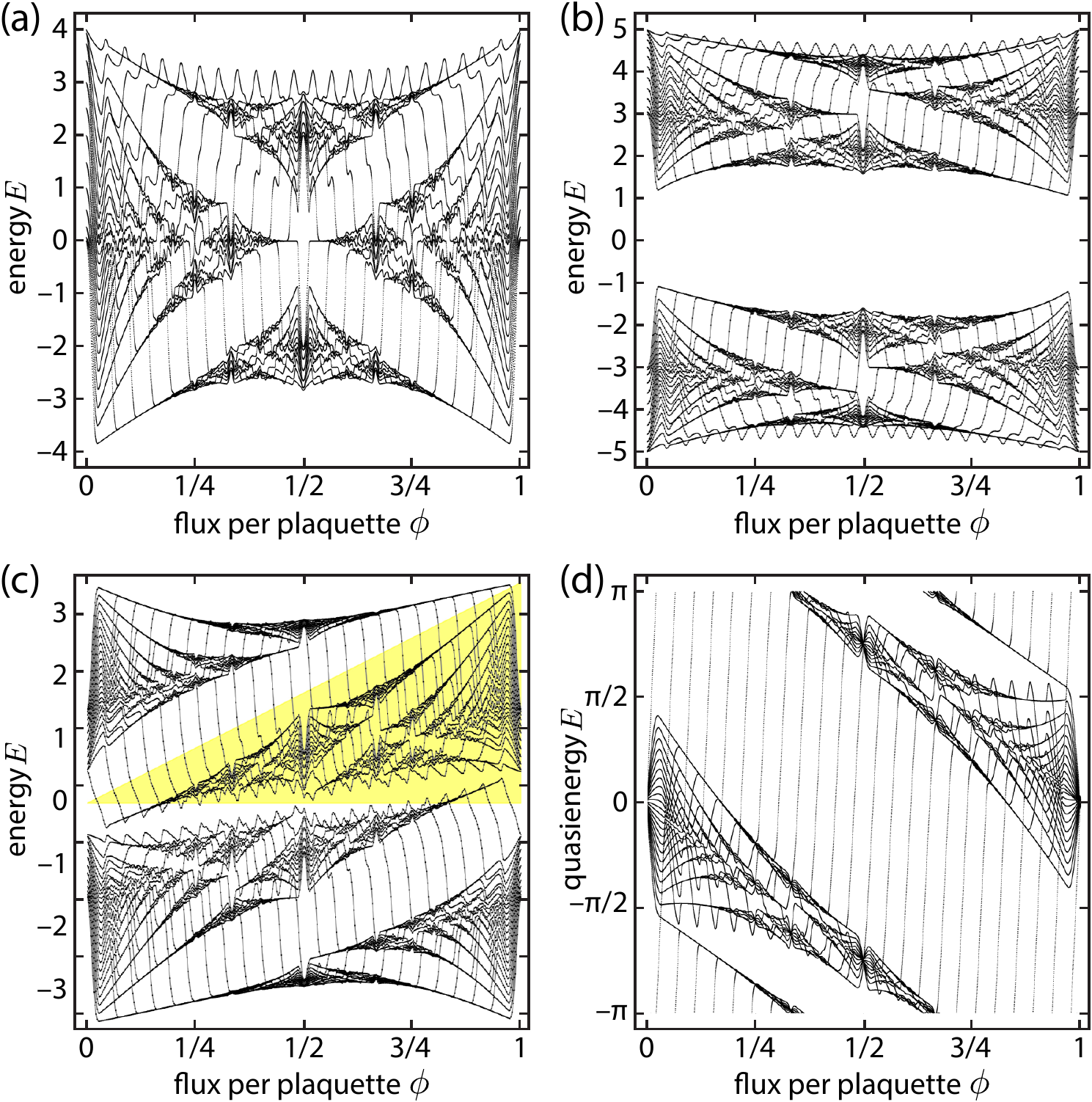}\caption{
Examples of Hofstadter spectra analogous to those shown in Fig.~\ref{fig:four_examples}, but computed for a lattice of finite width along $y$ (24 sites with periodic boundary conditions), and still infinite along $x$.
For better visibility of the spectral flow, the spectra are shown for a given quasimomentum only (\mbox{$k_x=0$}).}
\label{fig:four_examples_flow}
\end{figure}

All our results hinge on the fact that in a lattice of fixed width
$N_y$, tuning the magnetic flux from one commensurate value
to the next (by an increase in $\phi$ of $1/N_y$)
induces a \emph{spectral flow} of energy eigenvalues.
We will prove that the spectral flow across each gap is equal to the
topological invariant $\topind$ of the gap, i.e., the net number of edge
modes traversing this gap at an edge.
Figure~\ref{fig:four_examples_flow} presents four different examples
of Hofstadter butterfly spectra computed for a lattice of finite
width, which show the spectral flow of eigenvalues across the
gaps with a nonzero $\topind$.
Our result is consistent with the Streda
formula~\cite{streda1982theory} and the Wannier's Diophantine
equation~\cite{wannier1978result}.
However, it also applies to periodically driven systems, where the
spectral flow of quasienergy (rather than energy) eigenvalues is
considered.

\section{Time-independent Hamiltonians} We consider a
two-dimensional band insulator on a square lattice, where each site
can host $N$ internal states.
A perpendicular magnetic field $B$ couples to the motion of the particle
via Peierls phases, with the vector potential $A = (B\hspace{1pt}y,0,0)$ chosen in the Landau gauge.
The magnetic field is initially set to have a rational magnetic flux per plaquette, i.e., $\phi= p / q$, with $p$ and $q$ relative prime.

We restrict our lattice to a strip that is infinite along the $x$-axis
and has a finite width, $N_y$, along the $y$-axis with open boundary
conditions.
The sites are therefore labelled by position indices
$n_x,n_y\in\mathbb{N}$, with $1\le n_y \le N_y$.
We choose $N_y= m \hspace{1pt}q$ for some $m\in\mathbb{N}$.
This makes the initial value of the magnetic flux
\emph{commensurate} with the system size, in the sense that the width
$N_y$ incorporates an integer number of magnetic unit cells (each cell has a width $q$).
Due to our choice of the gauge and the geometry, the system is
translational invariant along $x$ for all values of $\phi$.
Hence, it is described by a single-particle Hamiltonian $\HH(k_x)$,
which is periodic in $k_x$ with period $2\pi$.
Its eigenvalues are the energies $E_j(k_x,\phi)$, with $j=1,\ldots,N
N_y$.

We shall focus on one of the energy gaps of the bulk Hamiltonian,
which we label by an energy value $\Egap$ well inside the gap.
Eigenstates of the system at $\Egap$, if they exist, are edge states,
with wavefunctions exponentially decaying towards the bulk;
the maximum decay length of these states is denoted by $\lambda$.
By choosing $m$ sufficiently large, we can safely assume $N_y \gg
\lambda$, so these states can be assigned to either the upper or
the lower edge.
Thus, edge modes, which are sections of the dispersion relation of
$\HH(k_x)$ which intersect $\Egap$,
can be assigned to either the
upper or lower edge depending on which edge their wavefunction at
energy $\tilde E$ belongs to.
We denote the edge mode energies by $\Eupper_r(k_x,\phi)$ and
$E^{\Lower}_s(k_x,\phi)$, for the upper and lower edge, respectively
($s$ and $r$ designate the index of the edge modes).
The topological invariant $\topind$ of the gap is the net number
of edge modes at the upper edge, with the right- and left-propagating edge
modes counted with opposite signs.

We study how the spectrum of the edge states depends on the magnetic
flux, as this flux is tuned from one commensurate value to the next.
We parametrize this process by $\beta\in[0,1]$, as 
\begin{align}
	\label{eq:parametrization_phi}
\phi = \frac{p}{q} + \frac{\beta}{N_y}, \quad \beta\in[0,1].
\end{align}
At the bottom edge, in a region of width $\lambda$, the change in
$\phi$ induces a change in the vector potential $A$ is of the order of
$\beta\lambda/N_y$, which vanishes in the limit \mbox{$N_y\to\infty$}.
Thus, bottom edge states are essentially unaffected.
At the top edge, in a region of width $\lambda$, however,
$A_x$ is increased approximately uniformly by
$\beta\hspace{1pt}\Phi_0$, up to corrections of the order of
$\lambda/N_y$.
As a result, the upper edge modes are cycled 
across the whole Brillouin zone,
\begin{align}
\label{eq:eupper_flow}
\Eupper_r(k_x,\phi)\approx \Eupper_r(k_x-2 \pi\beta,p/q), 
\end{align}
up to corrections of the order of $\lambda/N_y$.

We define the spectral flow across the gap as the net number of times
the energy eigenvalues of $\HH(k_x)$ cross the value $\Egap$, as
$\phi$ is tuned from $\phi = p/q$ to $\phi=p/q+1/N_y$, for
sufficiently large $N_y$ (so the bulk gap remains open).
Using the notation
\encapsulateMath{$\mathcal{F}_{p/q}^{p/q+1/N_y}(\tilde{E}; \{
  E_j(k_x,\phi) \},\phi)$} for this spectral flow, we have 
\begin{multline}
	\label{eq:spectral_flow_definition}
  \mathcal{F}_{p/q}^{p/q+1/N_y}(\tilde{E}; \{ E_j(k_x,\phi) \},\phi) \\
  = 
\int_{p/q}^{p/q+1/N_y} \!\!d\phi
\sum_{j} \frac{\partial E_j(k_x,\phi)}{\partial \phi} 
\delta(E_j(k_x,\phi)-\tilde{E}). 
\end{multline}
For a generic $k_x$, edge states at the lower edge gives no
contribution to this spectral flow since their spectrum is only
changed by $\mathcal{O}(1/N_y)$.
Bulk states also do not contribute to the spectral flow,
since the bulk gap remains open.
Hence, the spectral flow is given by the flow of the
upper edge states,
\begin{multline}
	\label{eq:spectral_flow_at_edge}
\mathcal{F}_{p/q}^{p/q+1/N_y}(\tilde{E}; \{ E_j(k_x,\phi) \},\phi)\\
	=
\mathcal{F}_{p/q}^{p/q+1/N_y}(\tilde{E}; \{ \Eupper_j(k_x,\phi) \},\phi).
\end{multline}
A nonzero spectral flow across a gap indicates that, as the magnetic
flux is tuned according to Eq.~(\ref{eq:parametrization_phi}), some
bulk states are transformed into upper edge states, are shifted in
energy across the gap, and eventually become bulk states again at the
end of the cycle.

\begin{figure}[b]
\includegraphics[width=\columnwidth]{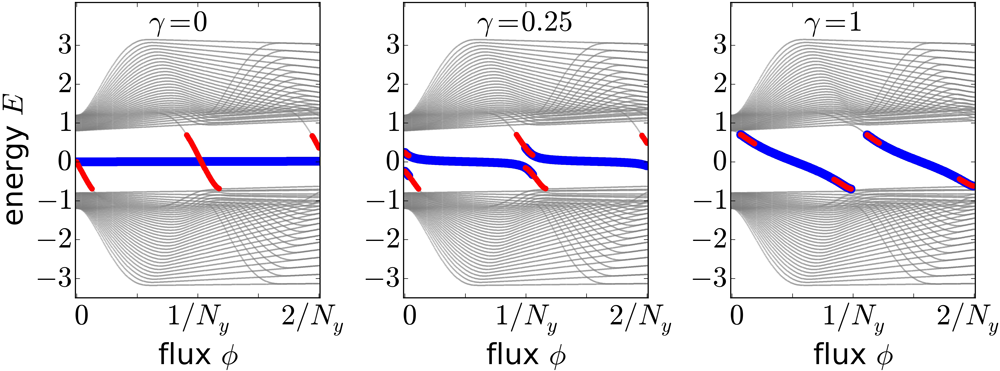}
\caption{Spectra of the Qi-Wu-Zhang model as a function of the magnetic flux (two cycles shown), computed on a lattice of finite width (36 sites) along $y$ and for a fixed quasimomentum ($k_x=0$).
    $\gamma$ indicates the hopping amplitude at the edge.
    ($\gamma=0$ for open, $\gamma=1$ for periodic boundary conditions).
        States with more than $\SI{30}{\percent}$ weight in the first two rows at the top (bottom) edge are marked by a thick red (blue) lines.
    \label{fig:qwz_B_kspectr}}
\end{figure}

Hence, we obtain the first result of this work: the spectral flow
across a gap is equal to the topological invariant $\topind$ of the
gap.
    This follows from Eqs.~\eqref{eq:eupper_flow} and
\eqref{eq:spectral_flow_at_edge}, which together give
\begin{multline}
  \mathcal{F}_{p/q}^{p/q+1/N_y}(\tilde{E}; \{ E_j(k_x,\phi) \},\phi) \\
  =   \mathcal{F}_{0}^{2\pi}(\tilde{E}; \{ \Eupper_j(k_x,\phi)\},k_x),
\label{eq:sigma_spectral_flow}
\end{multline}
where the quantity on the right hand side is the net number of edge states $E_r^{\Upper}(k_x,\phi)$ 
crossing the midgap energy $\tilde{E}$, as a function of $k_x$.
The latter is by definition the topological invariant $\topind$ of the gap.
This also proves that the spectral flow is independent of the choice of
the generic quasimomentum $k_x$.

Since our proof above relies on open boundary conditions, it is
worthwhile to show that the spectral flow is equal to $\topind$ also
in a system with periodic boundaries.
We therefore introduce an extra hopping amplitude, $\gamma$,
connecting opposite edges of the strip directly, with $0\leq
\gamma\leq 1$:
We thus obtain periodic boundary conditions with a single defect along
the stitching line instead of two separate edges.
Moreover, at $\gamma=1$, this defect line entirely disappears for any
commensurate value of $\phi$;
In these cases,
the spectrum is completely gapped around $\Egap$ for any value of
$k_x$.
Regardless of the value of $\gamma$, the spectral flow as defined in
Eq.~(\ref{eq:spectral_flow_definition}) is always an integer, since it
counts the number of states crossing $\tilde{E}$.
Bulk states do not contribute to it since their spectrum is
independent of $\gamma$, and the bulk gap stays open.
The only contribution thus arises from states that are localized near
the defect line.
These arise from the hybridization of the edge states due to the
nonzero value of $\gamma$, as shown with an example in
Fig.~\ref{fig:qwz_B_kspectr}.
However, as $\phi$ is increased to the next commensurate value, the
spectra of these midgap states must return to their initial values, up
to corrections of $\mathcal{O}(1/N_y)$.

\section{Global topology of the Hofstadter butterfly}  We now use
Eq.~(\ref{eq:sigma_spectral_flow}) to study the global topological
features of multiband Hofstadter butterflies. We address here the case
of time-independent Hamiltonians. Consider one of the bulk gaps of
the Hofstadter butterfly among those that stay open for all values of
the magnetic flux $\phi$.
At $\phi=0$, this corresponds to the $n_0$th gap, meaning that there
are $n_0$ bands with energy below it.
As $\phi$ is continuously tuned from $0$ to $1$, the gap must flow
into one of the gaps of the spectrum at $\phi=1$;
This is the $n_1$th gap at $\phi=1$.
We shall prove that the shift of the gap, $n_1-n_0$, is related to its
topological invariant as follows:
\begin{align}
  n_1-n_0 = \topind.
  \label{eq:shift_gap}
\end{align}

To prove Eq.~\eqref{eq:shift_gap}, we adopt the same
setting as above: an infinite strip of fixed width $N_y$ along the
$y$-axis, with a given quasimomentum $k_x$ along the $x$-axis.
Boundary conditions along the $y$-axis can be freely chosen to be,
e.g., periodic.
Moreover, $N_y$ has to be sufficiently large to have bulk states for
all $\phi$.
As $\phi$ is varied, we keep track of the gap by introducing a
continuous function $\Econt(\phi)$ taking midgap energies.

We prove Eq.~\eqref{eq:shift_gap} by showing that the number of states
in the spectrum at $\phi=1$, in the energy interval bounded by
$\Econt(0)$ and $\Econt(1)$, is given by $\nu$.
This number is equal to the net spectral flow of eigenvalues into the
energy region bounded by $\Econt(\phi)$ on the one side and by the
constant energy value $\Econt(0)$ on the other side, as indicated by
the highlighted region in the example in
Fig.~\figref{fig:four_examples_flow}{c}.
The net flow across the constant $\Econt(0)$ is zero, since the
\emph{total} spectral flow across any fixed energy value $\bar{E}$
must always vanish,
\begin{align}
	\label{eq:zeroTotalSpectralFlow}
\mathcal{F}_0^1(\bar{E};\{E_j\},\phi) &= 0,
\end{align}
a direct consequence of the periodicity of the spectrum in the
variable $\phi$.
The net flow across $\Econt(\phi)$, however, can be nonzero, as is the
case for a gap with a nontrivial topological invariant $\nu$.
In fact, decomposing the shift in $\phi$ from 0 to 1 into $N_y$ small
steps, Eq.~(\ref{eq:sigma_spectral_flow}) shows that
$N_y\hspace{1pt}\nu$ eigenvalues flow across $\Econt(\phi)$.
This means that at $\phi=1$, 
the interval between $\Econt(0)$ and $\Econt(1)$ contains $\abs{\nu}$
energy bands; moreover, the sign of the spectral flow $\nu$ of the
band tells us whether $\Econt(1)$ is larger or smaller than
$\Econt(0)$, thus concluding the proof of Eq.~(\ref{eq:shift_gap}).

\section{Floquet insulators} 
We next study the spectral flow and the Hofstadter butterfly in
periodically driven (i.e., Floquet) multiband insulators.
The Floquet insulator is a lattice Hamiltonian as discussed above,
with some of its parameters depending explicitly on time, periodically
with period $T$.
To define the Hofstadter butterfly, we include a
\emph{time-independent magnetic field} through Peierls phases, as for
the case of static Hamiltonians above.
Hence, the time-dependent Hamiltonian $\HH(\phi,\tau)$ is 
periodic both in time and in the magnetic flux,
\begin{align}
\HH(\phi,\tau)=\HH(\phi,\tau+1)=\HH(\phi+1,\tau),
\label{eq:periodic_driving}
\end{align}
where the quantity $\tau = t/T$ expresses time $t$ in dimensionless units.
This Hamiltonian defines the time evolution over one period of the
drive through the Floquet operator, \encapsulateMath{$\UU(\phi) =
  \mathcal{T}\hspace{-2pt} \exp[{-i T/\hbar \int_0^1 \hspace{-1pt}
      \HH(\phi,\tau)\hspace{1pt}
       \hspace{2pt} \mathrm{d}\tau}]$}, where
$\mathcal{T}\hspace{-2pt}$ denotes  time ordering.
The phases of the eigenvalues of the Floquet operator $\UU$ are the 
\emph{quasienergies} $\epsilon$,
which play the role of the
energy $E$ in a static Hamiltonian.
The quasienergies are defined to be in the interval $[-\pi, \pi]$,
with the endpoints $\epsilon=-\pi$ identified with $\epsilon=\pi$.
We call this interval the \emph{Floquet zone} of quasienergies, in
analogy to the Brillouin zone of quasimomenta.
As an example of Floquet Hofstadter butterfly,
Fig.~\figref{fig:four_examples}{d} shows the spectrum of quasienergies
as a function of the flux $\phi$ in the case of the Rudner
model of Ref.~\citenum{rudner2013anomalous}.

We will show how to adapt our results on time-independent Hamiltonians
to Floquet systems.
The result of Eq.~\eqref{eq:sigma_spectral_flow} directly carries over
to the quasienergies of a Floquet system:
The topological invariant of each gap is equal to the spectral
flow of quasienergies across it.
Further, the global topology of a Floquet Hofstadter butterfly can
also be related to the topological invariants of the gaps, in a
similar fashion as in Eq.~\eqref{eq:shift_gap}.
Concerning the latter, however, some further remarks are in order.

First, showing that the total spectral flow of quasienergies across a
constant $\bar{E}$ vanishes, see Eq.~(\ref{eq:zeroTotalSpectralFlow}),
is more demanding for the Floquet case than in the nondriven case.
To prove this,
we deform $\UU(\phi)$ continuously to $\hat{1}$
(the unity operator) by replacing $\HH(\phi,t)$ with
$\eta \hspace{1pt} \HH(\phi,t)$, where $\eta\in[0,1]$.
Since the total spectral flow across a fixed quasienergy
$\bar{\epsilon}$ is an integer-valued, continuous function of $\eta$,
its value must be independent of $\eta$.
At $\eta=0$, the total spectral flow vanishes because  
$\UU(\phi)=\hat{1}$, independently of $\phi$; thus, it also vanishes
at $\eta=1$.

Second, 
it is convenient to describe the spectral flow of quasienergies in a
scheme of repeated Floquet zones (in analogy to the repeated
Brillouin zones of quasimomentum).
Quasienergies can flow from one Floquet zone into a neighboring one,
as $\phi$ is tuned.
However, since the total spectral flow vanishes, they must return to
the original zone at $\phi=1$.
Hence, the same arguments used to prove Eq.~\eqref{eq:shift_gap} apply
in the repeated Brillouin zone scheme: gaps that stay open for all
values of $\phi$ must be shifted by $\nu$ energy bands as $\phi$
is tuned from 0 to 1.
Note that the difference $n_1-n_0$ is well defined in the repeated
Floquet zone scheme, even though $n_0$ and $n_1$ individually are not.
Once quasienergies are ``folded back'' into the first zone, the flow
of quasienergy gaps into neighboring Floquet zones can 
result in a winding of the whole Floquet Hofstadter butterfly in
quasienergy, as shown by the example in
Fig.~\figref{fig:four_examples_flow}{d}.

\section{Physical approach to the Rudner topological invariant}
The net number of edge states traversing the quasienergy gap at the
edge of the Floquet zone is a topological invariant that is unique to
periodically driven systems.
This number, the Rudner invariant $R$, modifies the bulk-edge
correspondence of the effective Hamiltonian
\cite{rudner2013anomalous}:
The net number of edge states traversing the $n$-th quasienergy gap is
given by $\topind_n = R+ \sum_{m<n} C_m$, where $C_m$ is the Chern
number of the $m$-th quasienergy band.
Unlike the Chern numbers, $R$ cannot be obtained from the bulk Floquet
operator $\UU(k_x,k_y)$ as a function of the quasimomenta $k_x,k_y$.
Rudner \emph{et al.} \cite{rudner2013anomalous} identified
$R$ with a winding number,
\begin{align}
  R &= \int_0^1\hspace{-2pt} \text{d}\tau \int_{\text{BZ}} \hspace{-2pt} \text{d}^2k  \hspace{2pt}\tr\hspace{-2pt}\left\{
  \hat{V}^\dagger \partial_\tau \hat{V}\hspace{1pt} [
  \hat{V}^\dagger \partial_{k_x} \hat{V},
  \hat{V}^\dagger \partial_{k_y} \hat{V}] \right\} \hspace{-2pt},
\label{eq:rudner_winding}
\end{align}
where $\hat{V}$ is the periodized Floquet operator, 
\begin{align}
  \hat{V}&(\tau,k_x,k_y)= e^{i\Heff \tau}         \mathcal{T}\hspace{-2pt} e^{-i\int_0^\tau \HH(\tau',k_x,k_y) \text{d}\tau'}.
\label{eq:def:periodized}
\end{align}

Based on the spectral flow, we take a direct physical approach to the
Rudner invariant, and obtain a simple formula for it, in contrast to
the abstract definition in Eqs.~\eqref{eq:rudner_winding} and
\eqref{eq:def:periodized}.
To show this, we consider the determinant of $\UU(\phi)$, which can be
expressed as 
\begin{align}
  \det \UU(\phi) &= \lim_{M\to\infty}
  \prod_{j=1}^M \exp\left[-i\hspace{2pt}\tr \HH(\phi,j/M)/M\right].
\end{align}
Each of the factors in the product on the right hand side is
independent of $\phi$, since the Peierls substitution only modifies
off-diagonal matrix elements of the instantaneous Hamiltonian.
It follows that $\det \UU(\phi)$ is independent of $\phi$, and
consequently that the sum of quasienergies, $\sum_j \epsilon_j(\phi)$,
can only increase or decrease as a function of $\phi$ in steps of
$2\pi$.
A step change in this sum happens whenever a quasienergy value
flows across boundary of the first Floquet zone, $\epsilon=\pm\pi$, in
the positive or negative direction.
Using the relation between spectral flow and the topological invariant of the gap, see Eq.~(\ref{eq:sigma_spectral_flow}), the net number of such crossings
as $\phi$ is tuned from one commensurate value to the next is given by
the topological invariant of the quasienergy gap comprising 
$\varepsilon=\pi$, i.e., the Rudner invariant $R$.
Thus, we obtain the Rudner invariant as    
\begin{align}
  R &= \frac{1}{2\pi}
  \Big\{ \sum_j \epsilon_j(k_x,\phi+1/N_y) - \sum_j \epsilon_j(k_x,\phi) \Big\},
\label{eq:rudner_differential}
\end{align}
where $k_x$ is chosen arbitrary.
{\setlength{\parskip}{1pt plus 0pt minus 0pt}
Our formula in Eq.~\eqref{eq:rudner_differential} is a direct route to
the Rudner invariant.
It shows that, although the bulk Floquet operator is not sufficient to
obtain the Rudner invariant, its evaluation at two commensurate values of the
magnetic flux is.
Moreover, Eq.~\eqref{eq:rudner_differential} might be more efficient
to compute than Eq.~\eqref{eq:rudner_winding}, as no numerous
derivatives of the Floquet operator are required.
}

\section{Conclusions}
We have shown that for lattice Hamiltonians, the spectral flow induced
by magnetic field across bulk gaps is a powerful physical concept.
It allows us to connect the topological invariant of the gap to the
global topology (connectedness) of the Hofstadter butterfly.
The concept of spectral flow also applies to periodically driven
systems, where it leads to a physically intuitive expression for 
the Rudner topological invariant.
Our theoretical work could find application in experiments
that aim to explore the Hofstadter butterfly or Floquet topological
phases using periodic driving~\cite{Groh:2016}.

\begin{acknowledgments}
We acknowledge financial support from the the ERC grant DQSIM and from the Deutsche Forschungsgemeinschaft SFB TR/185 OSCAR.
J.K.A. also acknowledges support from the Hungarian Scientific Research
Fund (OTKA) under Contract No. NN109651, and from the Janos Bolyai Scholarship of the Hungarian Academy of Sciences.
\end{acknowledgments}

\bibliographystyle{apsrev4-1}

\bibliography{walkbib}{}

\end{document}